\documentclass[prl,twocolumn]{revtex4}%
\usepackage{amsfonts}
\usepackage{amsmath}
\usepackage{amssymb}
\usepackage{graphicx}%
\begin{document}
\preprint{ }
\title{Charge-Inhomogeneity doping relations in YBa$_{2}$Cu$_{3}$O$_{y}$ detected by
Angle Dependent Nuclear Quadrupole Resonance}
\author{Rinat Ofer, Shahar Levy, Amit Kanigel, and Amit Keren}
\affiliation{Department of Physics, Technion - Israel Institute of Technology, Haifa 32000, Israel.}
\pacs{PACS number}

\begin{abstract}
The origin of charge inhomogeneity in YBa$_{2}$Cu$_{3}$O$_{y}$ is investigated
using a new experimental method designed to determine the nuclear quadrupole
resonance (NQR) asymmetry parameter $\eta$ for very wide NQR lines at
different positions on the line. The method is based on the measurement of the
echo intensity as a function of the angle between the radio frequency field
$\mathbf{H}_{1}$ and the principal axis of the electric field gradient. Static
charge inhomogeneity deduced from $\eta>0$ are found in this compound, but
\emph{only} in conjunction with oxygen deficiency. This limits considerably
the possible forms of charge inhomogeneity in bulk YBa$_{2}$Cu$_{3}$O$_{y}$.

\end{abstract}
\maketitle
\date{\today}

The discussion on the mechanism for high temperature superconductivity (HTSC)
is focused these days on the presence or absence of charge and spin
inhomogeneity in the CuO$_{2}$ planes. Such inhomogeneity can lead to a one
dimensional boundary, possibly in the form of stripes, between
\textquotedblleft hole-rich\textquotedblright\ and \textquotedblleft
hole-poor\textquotedblright\ regions and, allegedly, to
superconductivity.\cite{EmeryPRB97} Indeed, it is now established by both
surface and bulk techniques that most of the underdoped cuprates phase
separate into \textquotedblleft hole-rich\textquotedblright\ and
\textquotedblleft hole-poor\textquotedblright\ regions. For example, there is
a consensus derived from surface sensitive STM experiments \cite{LangNature02}
on underdoped Bi$_{2}$Sr$_{2}$CaCu$_{2}$O$_{8+\delta}$ that the planes are
inhomogeneous. A different example is La$_{2-x}$Sr$_{x}$Cu$_{1}$O$_{4}$ [LSCO]
where the evidence for phase separation is derived from experiments sensitive
to both magnetic fluctuation, such as muon spin relaxation ($\mu$SR)
\cite{NiedermayerPRL98}, and charge fluctuations such as nuclear quadrupole
resonance (NQR).\cite{SingerPRL02} Even in very underdoped YBa$_{2}$Cu$_{3}%
$O$_{y}$ (YBCO$_{y}$) phase separation is observed; neutron scattering from
phonons related to charge inhomogeneity is found in doping levels up to
YBCO$_{6.35}$ \cite{MookPRL02}, and $\mu$SR detects a spin glass phase for a
similar doping range.\cite{SannaPRL04} However, the origin of this phase
separation is still not clear; is it coming from competing phases in the
CuO$_{2}$ plane, or does it simply stem from the quenched disorder introduced
by doping?

A possible way to address this question is to apply NQR to YBCO$_{y}$ since in
this compound one can distinguish between different in-plane coppers [Cu(2)]
resonance lines, and associate each line with a local environment. However, a
standard NQR based determination of charge homogeneity is not available to
date due to the complicated and very wide spectrum of underdoped YBCO (which
will be reviewed below). The purpose of this work is to overcome the NQR
problems and to shed light on the evolution of charge homogeneity in the bulk
of YBCO close to optimal doping. For this purpose we developed a new
experimental technique based on the measurement of the Cu(2) NQR echo
intensity as a function of the angle between the radio frequency (RF) field,
$\mathbf{H}_{1}$, and the principal axis of the electric field gradient (EFG)
at the copper site. We call this angle-dependent NQR technique ADNQR.%

\begin{figure}
[h]
\begin{center}
\includegraphics[
height=3.1151in, width=4.3414in
]%
{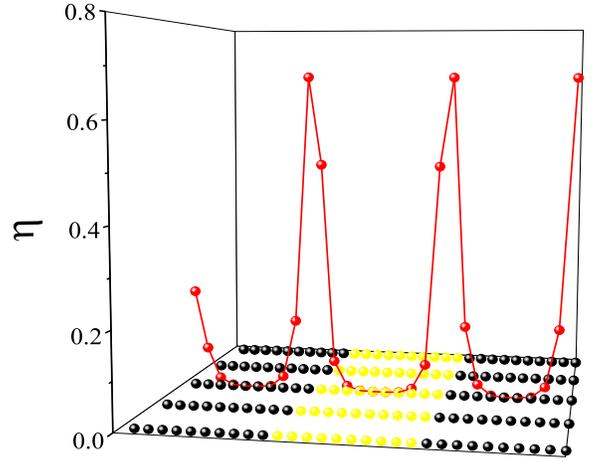}%
\caption{(Color online) A toy model calculation for $\left\vert \eta
\right\vert $ in the presence of charge stripes in the CuO$_{2}$ plane. See
text for details.}%
\label{toymodel}%
\end{center}
\end{figure}

Standard NQR is based on the fact that nuclei with spin $I>1/2$ can be viewed
as positively charged oval objects. As a result, their energy inside a solid
depends on their orientation in the electrostatic potential $V(\mathbf{r})$
generated by the other nuclear and electronic charges. When the nuclear poles
are close to positive charges their energy is high, and when the poles are
close to negative charges the energy is low. The energy difference between
different orientations is determined by the EFG tensor $V_{ij}=\frac
{\partial^{2}V}{\partial x_{i}\partial x_{j}}$ at the position of the Cu
nuclei. The directions can be chosen so that $V_{ij}$ is diagonal. These
directions are known as the principal axis of the EFG. Due to Laplace equation
($V_{xx}+V_{yy}+V_{zz}=0$) the NQR Hamiltonian is determined by only two
parameters, $\nu_{q}$ and $\eta$, and is given by
\begin{equation}
\mathcal{H}_{q}=\frac{\hbar\nu_{q}}{6}\left[  3I_{z}^{2}-\mathbf{I}^{2}%
+\eta\left(  I_{x}^{2}-I_{y}^{2}\right)  \right]  \label{Hamit}%
\end{equation}
where $\nu_{q}$ is a frequency scale proportional to the EFG $V_{zz}$ and
\begin{equation}
\eta=\frac{V_{yy}-V_{xx}}{V_{zz}} \label{etadef}%
\end{equation}
is a dimensionless number. It is customary to choose the directions so that
$\left\vert V_{xx}\right\vert \leq\left\vert V_{yy}\right\vert \leq\left\vert
V_{zz}\right\vert $, and therefore $0\leq\eta\leq1$. For the spin 3/2 Cu
nuclei this Hamiltonian has only one resonance frequency given by%
\begin{equation}
f=\nu_{q}\sqrt{1+\frac{\eta^{2}}{3}}. \label{resonance}%
\end{equation}

In YBCO$_{7}$ and YBCO$_{6}$ the directions are known experimentally:
$\widehat{\mathbf{z}}$ is the $\widehat{\mathbf{c}}$ direction and
$\widehat{\mathbf{x}}$ and $\widehat{\mathbf{y}}$ are directions in the
CuO$_{2}$ plane \cite{JPSJ1,SchwarzPRB90}. In YBCO$_{6.5}$ these directions
are determined to be the same on theoretical grounds \cite{SchwarzPRB90}, and
are believed not to vary for all other doping \cite{HaasePRB04}. Assuming the
directions are doping independent, $\eta$ is a measure of the local charge
homogeneity of the CuO$_{2}$ planes. When these planes are homogeneous with
local $xy$ rotation symmetry $\eta=0$. In contrast, when the planes are
inhomogeneous due to phase separation as in the case of static stripes, for
example, then on the boundary between hole poor and hole rich stripes the $xy$
rotation symmetry will be lost and we expect $\eta\neq0$. This situation is
demonstrated in Fig.~\ref{toymodel} obtained using a toy model. The parameters
of this model are chosen for clarity only and have no implications on the
physical situation in YBCO. A plane of a square lattice with total charge $Q$
is sandwiched between two similar planes with total charge $-Q/2$ each. The
electric field produced by each ion is screened with a screening length of two
lattice sites. The charge in the central plane is distributed in the form of
stripes, 10 atoms wide, and $\eta$ is numerically calculated. In the figure
only the central plane with its stripes, and $\left\vert \eta\right\vert $
along one line of ions crossing different stripes, are shown. This figure
demonstrates that as a result of the stripes $\eta\neq0$ for all nuclei on the
boundary between stripes, despite the fact that the underlying lattice is
square. It should be strongly emphasized, however, that we use a stripes-based
toy model for the demonstration because of its simplicity, and our experiment
has no bearing on the model developed in Ref.~\cite{KivelsonRMP03} where the
stripes are dynamic on a time scale much shorter than our experimental one.

\begin{figure}
[h]
\begin{center}
\includegraphics[
height=3.0502in,
width=3.8156in
]%
{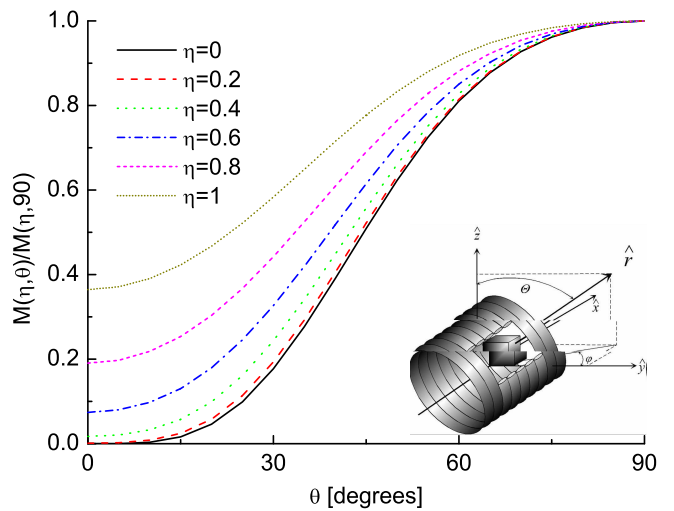}%
\caption{{(Color online) The expected echo intensity for various values of
$\eta$ in an oriented powder (averaged azimuthal angle $\phi$) as a function
of the polar angle $\theta$ between the RF field }$\mathbf{H}_{{1}}${ and the
}$\widehat{\mathbf{z}}${ direction of the EFG. The pulse length is optimized
at $\theta=90$ and kept constant throughout the sample rotation. The inset
shows the experimental configuration.}}%
\label{powder}%
\end{center}
\end{figure}

Our objective in this work is to determine $\eta$ as a function of doping in
different environments using ADNQR, which is described in detail elsewhere
\cite{LevyJMR}. In this modification of standard NQR one measures the echo
intensity in a coil, whose symmetry axis is in the direction
\[
\widehat{\mathbf{r}}=(\sin\theta\cos\phi,\sin\theta\sin\phi,\cos\theta)
\]
with respect to the principal axis of the EFG, as a function of $\theta$ and
$\varphi$. This situation is demonstrated in the inset of Fig.~\ref{powder}.
The echo is obtained using a $\pi/2-\tau-\pi$ pulse sequence. The pulses are
of frequency $f$, duration $t_{\pi/2}$ and $2t_{\pi/2}$, and create a field
$H_{1}=\omega_{1}/\gamma$, where $\gamma$ is the copper nuclei gyromagnetic
ratio. The basic idea behind this experiment is that when $\widehat
{\mathbf{r}}$ is in the $\widehat{\mathbf{z}}$ direction, and $\eta=0$, no
spin transitions will occur and no echo will be formed. In contrast when
$\eta\neq0$ an echo will be formed even though $\widehat{\mathbf{r}}$ is
parallel to $\widehat{\mathbf{z}}$. For an arbitrary $\theta$ and $\varphi$
the nuclear magnetization $M$ in the coil, at the time of the echo, is given
\cite{PrattMolPhy77,LevyJMR} by
\begin{equation}
M\left(  \eta,\theta,\phi\right)  =M_{0}\lambda\sin^{3}(\lambda\omega
_{1}t_{\frac{\pi}{2}}) \label{MainResult}%
\end{equation}
where
\begin{equation}
\lambda\mathbf{=}\frac{1}{2\sqrt{3+\eta^{2}}}\left\{  \left[  9+\eta^{2}%
+6\eta\cos(2\phi)\right]  \sin^{2}(\theta)+[2\eta\cos(\theta)]^{2}\right\}
^{1/2}, \label{epsilon}%
\end{equation}
and $M_{0}$ is a constant.%

We apply the ADNQR experiment to oriented powder of YBa$_{2}$Cu$_{3}$O$_{y}$
with $\widehat{\mathbf{c}}||\widehat{\mathbf{z}}$. In such powders the
$\widehat{\mathbf{a}}$ and $\widehat{\mathbf{b}}$ directions are mixed. In
this case $M$ is obtained from Eq.~\ref{MainResult} by averaging over $\phi$,
namely,
\begin{equation}
M\left(  \eta,\theta\right)  =\frac{1}{2\pi}\int\limits_{0}^{2\pi}%
\mathbf{M}\left(  \eta,\theta,\phi\right)  d\phi. \label{magnetization}%
\end{equation}
This averaging must be done numerically, and the expected $M\left(
\eta,\theta\right)  /M\left(  \eta,90\right)  $ for various values of $\eta$
is presented in Fig \ref{powder}. The pulse length ($t_{\pi/2}$) is optimized
at $\theta=90$ and is kept constant throughout the rotation of the sample with
respect to the coil.

The ADNQR technique has three major strengths: (I) It allows the determination
of $\eta$ even for a wide NQR spectrum and at every point on the spectrum,
thus providing a great advantage over nuclear magnetic resonance (NMR), which
can determine $\eta$ only from the entire spectrum with no local resolution;
(II) unlike NMR, where one needs to fit the spectrum to 5 different parameters
out of which only one is $\eta$, ADNQR is sensitive only to $\eta$; (III) it
allows the determination of $\eta$ without the application of a static
magnetic field and can be used even in the superconducting state. The weakness
of ADNQR is that it is very insensitive at small $\eta$'s as demonstrated in
Fig \ref{powder}.%

\begin{figure}
[h]
\begin{center}
\includegraphics[
height=3.7126in,
width=2.9551in
]%
{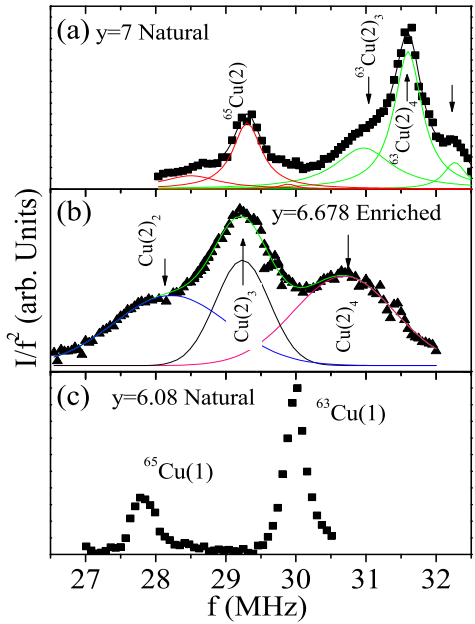}%
\caption{{(Color online) }NQR intensities versus frequency $f$ normalized by
$f^{2}$ in three different YBCO$_{y}$ samples. The YBCO$_{6.678}$ sample is
fully enriched with $^{63}$Cu. YBCO$_{7}$ and YBCO$_{6.08}$ have natural
abundance of $^{65}$Cu and $^{63}$Cu. Also shown are Gaussian fits indicating
the contribution to the spectrum of different Cu(2) environments. The arrows
show the frequencies where ADNQR is applied.}%
\label{Lines}%
\end{center}
\end{figure}

Before presenting the ADNQR results it is essential to review the NQR
frequency assignments in YBCO. In the (a), (b), and (c) panels of
Fig.~\ref{Lines} we show the NQR lines at a temperature of $100$~K for three
different samples with $y=7.0$, $6.678$, $6.08$, respectively. The $y=6.678$
sample is fully enriched with $^{63}$Cu. The lines are obtained by a
spectrometer with a home-made automated frequency sweep. The echo intensity
$I$ is normalized by $f^{2}$, where $f$ is the applied frequency, in order to
correct for population difference and the induced signal in the coil. The
solid lines in Fig.~\ref{Lines} are fits of Gaussians to the data indicating
that the spectrum is constructed from contributions of different local environments.

It is simplest to review the enriched $y=6.678$ spectrum in panel (b) first.
This spectrum exhibit three peaks at $f=28.1$, $29.2$, and $30.8$ MHz (none of
which are from the $^{65}$Cu isotope). The peaks are classified in terms of
the number of oxygen surrounding the chain copper [Cu(1)] neighboring the
detected Cu(2). The higher the oxygen coordination of the this Cu(1), the
higher the frequency. In other words, the peak in $f=30.8$~MHz is from Cu(2)
adjacent to a Cu(1) in a fully oxygenated environment [Cu(2)$_{4}$], the peak
in $f=28.1$~MHz is from Cu(2) nearing a Cu(1) in a fully oxygen deficient
environment [Cu(2)$_{2}$], and finally, the peak in $f=29.2$~MHz is from Cu(2)
whose neighboring Cu(1) is missing one oxygen [Cu(2)$_{3}$].

The main features in the spectrum of the natural YBCO$_{7}$ (without
enrichment) depicted in panel (a) are two resonance peaks at $f=31.55${}~MHz
and $f=29.3$~MHz. In addition, each peak has two shoulders to the left
($31$~MHz/$28.7$~MHz) and right ($32$~MHz/$29.7$~MHz) of the biggest peak. The
origin of the right shoulder is not clear. The left shoulder is associated
again with a partially oxygen deficient environment (mostly Cu(2)$_{3}$). This
spectrum, including the shoulders, resembles the spectrum obtained by others
\cite{VegaPRB89,YasuokaPT89}, where the lines are identified as the
plane-copper isotopes [$^{63}$Cu(2)] and [$^{65}$Cu(2)]. The ratio between the
peak intensity of the two isotopes is the one expected from their natural
abundance, indicating that our spectrometer functions properly.

It is important to mention that the Cu(1) can also contribute intensity to the
Cu(2) signal \cite{YasuokaPT89}. This is demonstrated in panel (c) for
YBCO$_{6.08}$. In this antiferromagnetic sample the Cu(2) signal is at
$f=90~$MHz \cite{JPSJ2} and in panel (c) only the two isotopes of Cu(1) can be
seen \cite{YasuokaPT89}. This panel clarifies our careful choice of
oxygenation level of the underdoped compound. The Cu(2) peaks of the $y=6.678$
sample are not contaminated by the Cu(1) signal.

The ADNQR experiments in the YBCO$_{7}$ sample were done on the $^{63}$Cu(2)
peak and the two shoulders. In YBCO$_{6.678}$ we investigated the peaks of all
different local environments. The frequencies where ADNQR was applied are
marked by arrows in panels (a) and (b). The ADNQR experiments were also done
at $T=100$~K temperature using an automated sample rotor. To improve $H_{1}$
homogeneity we used a spherical coil.

\begin{figure}
[h]
\begin{center}
\includegraphics[
height=3.8086in,
width=3.1375in
]%
{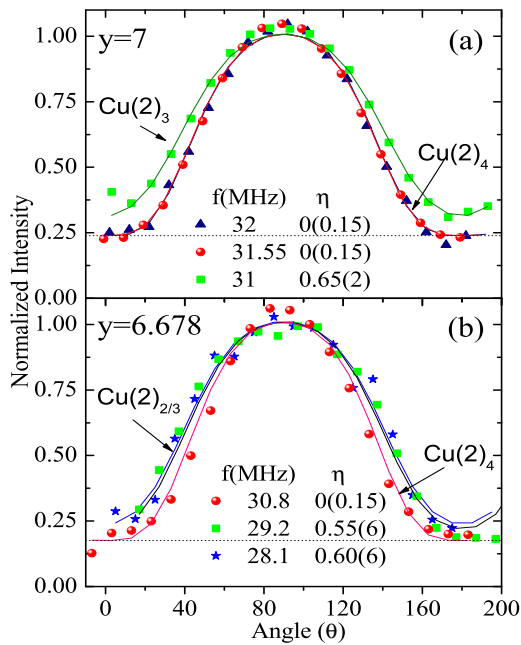}%
\caption{{(Color online) }The echo intensity as a function of the angle
$\theta$ between the RF field $\mathbf{H}_{1}$ and the principal axis of the
electric field gradient $\widehat{\mathbf{z}}$ for the two YBCO samples with
different oxygen level $y$, and at three different points in the specturm
shown in Fig.~\ref{Lines}(a) and (b). The solid lines are fits to
Eq.~\ref{magnetization} as described in the text.}%
\label{angledep}%
\end{center}
\end{figure}

The ADNQR results for YBCO$_{7}$ and YBCO$_{6.678}$ in the $\theta=0$ to $180$
range are depicted in Fig. \ref{angledep}. In all cases the intensities at
$\theta=0$ and $180$ are lower than at $\theta=90$, as it should be. It is
also clear that the intensity as a function of angular deviation from
$\theta=90$ drops faster for the Cu(2)$_{4}$ in both dopings. This is a
model-independent observation of the fact that $\eta$ is smallest for
Cu(2)$_{4}$. The fit of Eq.~\ref{magnetization} to the experimental data is
demonstrated by the solid lines. In the fit we allow a common finite base line
for a given sample, in order to account for some unknown amount of
misalignment. We found $24$\% and $17$\% misaligned crystallines in the
YBCO$_{7}$ and YBCO$_{6.678}$, respectively. This misalignment is typical. The
best fit for YBCO$_{7}$ peak gives $\eta=0\pm0.15$ in agreement with values
previously obtained by NMR for optimally doped YBCO, where the values $\eta=0$
\cite{PenningtonPRB88} and $\eta=0.01\pm0.01$ \cite{JPSJ1} were found. This
result reassures us that the experimental method and analysis are correct.
Surprisingly, our fit also gives $\eta=0\pm0.15$ for the Cu(2)$_{4}$ in
YBCO$_{6.678}$ sample ($f=30.8$ MHz). In contrast, in oxygen deficient
environments in both YBCO$_{7}$ and YBCO$_{6.678}$ $\eta\simeq0.6$. This is
the major finding of our work.%

The data from YBCO$_{6.678}$ are the first evidence for charge anisotropy in
such highly doped YBCO. As mentioned before, to date, static charge/spin
inhomogeneity has been detected in YBCO by neutron scattering and $\mu$SR only
up to $y\leq6.35$ \cite{MookPRL02,SannaPRL04}. However, the possibility of
electronic phase separation, reflected by $\eta>0$, is found (within our
experimental sensitivity) only in conjunction with quenched disorder. This
means that Cu(2)$_{4}$ cannot reside on a static boundary between two phases
of different charge concentration, like the boundary generated by our toy
model. This result is in accordance with recent STM measurements
\cite{McElroyScience05}. We thus conclude that in YBCO charge inhomogeneity
stems from quenched disorder.

We would like to thank H.~Alloul, P.~Mendels, and J.~Bobroff for very helpful
discussions. This work was funded by the Israeli Science Foundation and the
Posnansky Research Fund in High Temperature superconductivity.

\end{document}